\documentstyle [12pt,twoside]{article}

\newcommand{\bq}{\begin{equation}}
\newcommand{\ba}{\begin{eqnarray}}
\newcommand{\eq}{\end{equation}}
\newcommand{\ea}{\end{eqnarray}}

\def\b{\beta}

\def\p{\pi}

\def\bo{{\raise.15ex\hbox{\large$\Box$}}}
\def\bob{{\lower.2ex\hbox{\large$\Box$}}}
\def\pa{\partial}

\def\TH{{\raise.2ex\hbox{$\displaystyle \bigodot$}\mskip-4.7mu \llap H \;}}
\def\face{{\raise.2ex\hbox{$\displaystyle \bigodot$}\mskip-2.2mu \llap {$\ddot
        \smile$}}}

\def\Hat#1{\rlap{\kern.10em$\widehat{\phantom G}$}#1}
\def\HAt#1{\rlap{\kern.05em$\widehat{\phantom G}$}#1}

\def\cap#1{\rlap{\kern.1em$\widehat{\phantom{G\vrule height.8em}}$}#1{}}
\def\Cap#1{\rlap{\kern.05em$\widehat{\phantom{G\vrule height.8em}}$}#1{}}

\def\leftrightarrowfill{$\mathsurround=0pt \mathord\leftarrow \mkern-6mu
        \cleaders\hbox{$\mkern-2mu \mathord- \mkern-2mu$}\hfill
        \mkern-6mu \mathord\rightarrow$}
\def\overleftrightarrow#1{\vbox{\ialign{##\crcr
        \leftrightarrowfill\crcr\noalign{\kern-1pt\nointerlineskip}
        $\hfil\displaystyle{#1}\hfil$\crcr}}}

\def\frac#1#2{{\textstyle{#1\over\vphantom2\smash{\raise.20ex
        \hbox{$\scriptstyle{#2}$}}}}}

\catcode`@=11
\def\underline#1{\relax\ifmmode\@@underline#1\else
        $\@@underline{\hbox{#1}}$\relax\fi}
\catcode`@=12

\def\nis{\nointerlineskip}
\def\Abar{\vbox{\nis\moveright.33em\vbox{
        \hrule width.35em height.04em}\nis\kern.05em\hbox{$A$}}{}}
\def\Dbar{\vbox{\nis\moveright.20em\vbox{
        \hrule width.50em height.04em}\nis\kern.05em\hbox{$D$}}{}}
\def\Gbar{\vbox{\nis\moveright.20em\vbox{
        \hrule width.50em height.04em}\nis\kern.05em\hbox{$G$}}{}}
\def\mbar{\vbox{\nis\moveright.15em\vbox{
        \hrule width.60em height.04em}\nis\kern.05em\hbox{$m$}}{}}
\def\Rbar{\vbox{\nis\moveright.20em\vbox{
        \hrule width.50em height.04em}\nis\kern.05em\hbox{$R$}}{}}
\def\Vbar{\vbox{\nis\moveright.05em\vbox{
        \hrule width.60em height.04em}\nis\kern.05em\hbox{$V$}}{}}
\def\Xbar{\vbox{\nis\moveright.20em\vbox{
        \hrule width.60em height.04em}\nis\kern.05em\hbox{$X$}}{}}
\def\thetabar{\vbox{\nis\moveright.15em\vbox{
        \hrule width.30em height.04em}\nis\kern.05em\hbox{$\theta$}}{}}
\def\Lambdabar{\vbox{\nis\moveright.25em\vbox{
        \hrule width.35em height.04em}\nis\kern.05em\hbox{${\mit\Lambda}$}}{}}
\def\Sigmabar{\vbox{\nis\moveright.25em\vbox{
        \hrule width.50em height.04em}\nis\kern.05em\hbox{${\mit\Sigma}$}}{}}
\def\phibar{\vbox{\nis\moveright.18em\vbox{
        \hrule width.40em height.04em}\nis\kern.05em\hbox{$\phi$}}{}}
\def\chibar{\vbox{\nis\moveright.12em\vbox{
        \hrule width.40em height.04em}\nis\kern.05em\hbox{$\chi$}}{}}
\def\psibar{\vbox{\nis\moveright.23em\vbox{
        \hrule width.40em height.04em}\nis\kern.05em\hbox{$\psi$}}{}}
\def\debar{\vbox{\nis\moveright.18em\vbox{
        \hrule width.35em height.04em}\nis\kern.05em\hbox{$\partial$}}{}}
\def\delbar{\vbox{\nis\moveright.10em\vbox{
        \hrule width.63em height.04em}\nis\kern.05em\hbox{$\nabla$}}{}}

\begin{document}
\centerline{\large{\bf {Statistical Mechanics of Kinks in
(1+1)-Dimensions}}}

\vspace{1.5cm}

\centerline{Francis J. Alexander$^{\star}$ and  Salman
Habib$^{\dagger}$}

\vspace{1.5cm}

\centerline{\em $^{\star}$Center for Nonlinear Studies}
\centerline{\em and}
\centerline{\em $^{\dagger}$T-6, Theoretical Division}
\centerline{\em Los Alamos National Laboratory}
\centerline{\em Los Alamos, NM 87545}

\vspace{2.5cm}

\centerline{\bf Abstract}

We investigate the thermal equilibrium properties of kinks in a
classical $\phi^4$ field theory in $1+1$ dimensions. The distribution
function, kink density, and correlation function are determined from
large scale simulations. A dilute gas description of kinks is shown to
be valid below a characteristic temperature. A double Gaussian
approximation to evaluate the eigenvalues of the transfer operator
enables us to extend the theoretical analysis to higher temperatures
where the dilute gas approximation fails. This approach accurately
predicts the temperature at which the kink description breaks down.

\vfill
\noindent e-mail:\\
\noindent fja@goshawk.lanl.gov\\
\noindent habib@eagle.lanl.gov\\
\newpage

The statistical mechanics of coherent structures such as solitons and
solitary waves in nonlinear systems has been a subject of study for
some time \cite{SS}. Recent interest has been fueled by new
applications not only in condensed matter physics \cite{CM}\cite{BKT},
but also by potential applications in particle physics (sphalerons)
\cite{SR} and cosmology (domain walls, baryogenesis) \cite{KT}. In
this paper we focus on the classical equilibrium statistical mechanics
of solitary wave solutions (``kinks'') of a tachyonic mass $\phi^4$
field theory in $1+1$ spacetime dimensions with Lagrangian
\begin{eqnarray}
L= {1\over 2}\left(\pa_t\Phi\right)^2
   -{1\over 2}\left(\pa_x\Phi\right)^2
   +{1\over 2}m^2\Phi^2-{1\over 4}\lambda\Phi^4.                \label{1}
\end{eqnarray}
For our simulations we use the dimensionless form of this theory,
given by the transformations: $\Phi\to a\phi$, $x\to x/m$, and $t\to
t/m$, where $a^2=m^2/\lambda$. The equation of motion then becomes
\begin{eqnarray}
\pa^2_{tt}\phi=\pa^2_{xx}\phi -\phi\left(\phi^2-1\right).          \label{2}
\end{eqnarray}
What makes this model so useful is that its behavior is representative
of a large class of soliton-bearing systems.  Moreover, it is amenable
to both theoretical analysis and numerical simulation.

The statistical mechanics of kinks in this system has been studied by
two approaches. In the first, and phenomenological, approach one assumes
that the kinks and the fluctuations (``phonons'') about the asymptotic
field minima may be treated as weakly interacting elementary
excitations. The canonical partition function can then be found by
standard methods \cite{SS}\cite{KS}\cite{CKBT}. Alternatively, as
shown by Krumhansl and Schrieffer (KS) \cite{KS}, building on earlier
work of Sears, Scalapino, and Ferrell \cite{SSF}, it is possible to
calculate the partition function, in principle exactly, by exploiting
a transfer operator technique. KS showed that in the low temperature
(``dilute gas'') limit the partition function factorizes into a
contribution from a harmonic term and from a tunneling term which they
were able to identify with the phonon and kink contributions
respectively in the phenomenological theory.  The ideas of KS were
further refined and extended to a wider class of systems by Currie,
Krumhansl, Bishop, and Trullinger \cite{CKBT}. In particular,
interactions of kinks with linearized phonons were considered, leading
to substantial corrections to the results of KS.

A key result of these efforts is the prediction that the spatial
density of kinks
\begin{equation}
n_k\propto\sqrt{E_k\beta}\exp(-E_k\beta),                    \label{dgnum}
\end{equation}
where $E_k=\sqrt{8/9}m^3/\lambda$  ($E_k=\sqrt{8/9}$ for the
dimensionless form of the theory) is the kink energy, and $\beta$, the
inverse temperature (for the dimensionless case,
$\beta\to\beta/(a^3\sqrt{\lambda}$)). A related prediction is that at
low temperatures the field correlation length $\lambda$ defined by
\begin{equation}
\langle\phi(0)\phi(x)\rangle\sim\exp\left(-x/\lambda\right) \label{cl}
\end{equation}
has an exponential temperature dependence \cite{CKBT},
\bq
\lambda={1\over 4}\sqrt{{\p\over 3}}
{1\over\sqrt{E_k\b}}\exp(E_k\beta).      \label{CL}
\eq
Computer simulations to verify these results date back to Koehler,
Bishop, Krumhansl, and Schrieffer \cite{KBKS} who found only a
qualitative agreement. Recent work \cite{GR}\cite{BD}\cite{AFG} has
led to more detailed comparisons, however significant discrepancies
have been found over the temperature range explored in these
simulations. It has been speculated that these discrepancies are due
to finite size effects and phonon dressing of the bare kink energy
(breather contributions to the free energy may also be significant
\cite{SSB}). In this paper we show that the difficulty is partly that
previous simulations were not carried out at low enough temperatures.
Indeed, for the range of temperatures studied in Refs.
\cite{GR}\cite{BD}\cite{AFG}, the dilute gas approach is simply not a
valid approximation (there is also a problem with the operational
definition of kinks used by these authors over the range of
temperatures they studied).

We have studied the equilibrium statistical mechanics of kinks in
the $\phi^4$ model (\ref{1}) by implementing a Langevin code on a
massively parallel computer. The key idea is to supplement the
equation of motion (\ref{2}) with noise and viscosity terms obeying an
appropriate fluctuation-dissipation theorem so that the system is
driven to thermal equilibrium at the desired temperature \cite{ScS}. To
understand our results in the high and intermediate temperature region
not susceptible to a dilute gas analysis, we have used a double
Gaussian wave function approximation for the quantum
mechanical problem which results from applying the transfer operator
method. Our results are: (1) the dilute gas predictions for the kink
density and the correlation length are very accurate below a certain
(theoretically estimable) temperature, (2) above this temperature the
double Gaussian results for the kink number and the correlation length
agree with the simulations, (3) kinks are found to disappear above a
characteristic temperature, in good agreement with our theoretical
prediction, and (4) our Gaussian approximation accurately describes
the classical single point field distribution function at high and
intermediate temperatures where the dilute gas (WKB) approximation
breaks down.  We also observe a temperature dependence of the
asymptotic field value away from a kink which may be attributed to
kink-kink interactions.

The canonical partition function for the Lagrangian (\ref{1}) is
given by the functional integral
\begin{eqnarray}
Z = \int D\phi D\pi~\exp\left(-\beta H(\phi,\pi)\right) \label{Z}
\end{eqnarray}
where $\pi$ is the canonical momentum of the field and $H$, the field
Hamiltonian. The transfer operator technique \cite{SSF} reduces the
calculation of the partition function in the thermodynamic limit to
simply finding the ground state energy of the double well quantum
Hamiltonian (here written for the dimensionless case),
\begin{equation}
H_Q={1\over 2}p^2 - {1\over 2\beta^2}{\bar{\phi}}^2 + {1\over
4\beta^4}{\bar{\phi}}^4               \label{ham}
\end{equation}
where ${\bar{\phi}}=\beta\phi$. At low temperatures the two wells are
widely separated and the ground state energy is given by the
oscillator ground state energy for one of the wells minus the
tunnel-splitting term, usually calculated by WKB methods. The dilute
gas/WKB approximation for the kink number is valid when the
tunnel-splitting is small enough such that only the first two energy
eigenstates are necessary to
estimate the ground state energy of the Hamiltonian
(\ref{ham})\cite{BK}. At higher temperatures where kinks still exist,
higher energy states cannot be ignored. Since kinks are associated
with tunneling, we expect them to vanish when the ground state energy
is higher than the classical barrier height: this intuition is
confirmed by our simulations.

One can compare the simulations of the kink system with
numerical solutions for the energy eigenvalues of the Hamiltonian
$H_Q$. Instead, we take a different approach by implementing a
double Gaussian variational method (see \cite{AHK} for details)
which is an order of magnitude more accurate than the simple Gaussian
approximation \cite{GA} for this problem and correctly accounts for the
reduction of energy due to overlap terms in the wave functions of the
two wells, at least for moderate to large overlaps. Three
qualitatively different regimes exist: (1) all the energy eigenvalues
lie above the classical barrier, (2) the ground state energy lies
below the classical barrier height, and (3) the energy difference
between the ground and first excited state becomes negligible in
comparison with the energy difference between the ground and the
second excited state (this occurs at $\beta\sim 6.7$). Our simulations
confirm the theoretical expectations that there are no kinks in region
(1), that there are kinks, but that the dilute gas approximation is
invalid in region (2), and finally, that the dilute gas approximation
is accurate in region (3) (a regime unexplored in detail by previous
simulations).

We measure the classical single point field distribution function
$P[{\phi}]$ directly in our simulations. For the analogous quantum
mechanical problem this is just the square of the ground state wave
function $\Psi_0$. Results from the simulations and our theory are
compared in Fig. I and are in reasonable agreement. The presence of
kinks implies a double peak in $P[\phi]$ \cite{ADB}(the converse is
false) while a single peak at the origin means that kinks and thermal
phonons can no longer be distinguished. From the simulations such a
transition occurs at $\beta\simeq 1.7$, in agreement with the
theoretical calculation of when $\Psi_0^{~2}$ goes over from a double
to single peaked distribution. As expected, this is also the
temperature ($\beta=1.734$) where the ground state energy crosses the
classical barrier height (a discussion of various methods to determine
the characteristic temperature is given in Ref. \cite{ARB}). The
double peaks in the distribution function move inward from the
classical minimum as the temperature increases (this is clearly seen
in Fig. I). We attribute this to kink-kink interactions with the kinks
presumably sacrificing some potential energy to lower their strain
energy.

The Langevin equation for the dimensionless theory is
\begin{eqnarray}
\pa^2_{tt}\phi=\pa^2_{xx}\phi-\eta\pa_t\phi-\phi(1-\phi^2) + F(x,t).
\end{eqnarray}
To guarantee an approach to equilibrium, the Gaussian, white noise $F$
and the viscosity $\eta$ are related via the fluctuation-dissipation
theorem:
\begin{eqnarray}
\langle F (x,t) F (x',t') \rangle = 2\eta\beta^{-1}\delta (x-x')
\delta (t-t').
\end{eqnarray}

We carried out numerical simulations on lattices with 16384 sites and
solved the Langevin equation using a standard leapfrog algorithm.  The
time step was $\Delta t =0.02$ and the lattice spacing was $\Delta x =
0.5$.  For all of the simulation results reported here we used a
viscosity of $\eta =1.0$.

Our system size is one to two orders of magnitude larger than that in
most previous simulations. Large system sizes are necessary to get
acceptable statistics at low temperatures. Systems were evolved from
a random initial condition to equilibrium. The length of time
necessary to ensure equilibrium increased with inverse temperature.
For $\beta = 8$ the time required was approximately $10^7$ time steps,
and for the highest temperatures, less than $10^5$ steps.

Two quantities of interest reported here are the kink number and the
field correlation length.  To compute the kink number, we need an
operational way to identify kinks, although there is an exact kink
solution available theoretically. As a result we examine several
possible definitions, all of which rely on a knowledge of the kink
size.  From the classical solution for a kink centered at $x_0$,
$\phi=tanh((x-x_0)/\sqrt{2})$, the kink scale $L_k$ is approximately 8
lattice units.  Raw kink configurations are shown in Fig. II. At low
temperatures ($\beta >6$), kinks may be identified easily, however at
higher temperatures this is clearly not the case.

The simplest thing to do is to count the number of zero-crossings of
the field, since one may argue that these are the ``tunneling events''
which correspond to kinks. However, at higher temperatures there are
zero-crossings due to thermal noise, and counting all zero-crossings
would lead to a gross overestimation of the number of kinks. A
possible solution is to use a smoothed field by either averaging or
block-spinning the actual field configuration over a length of the
order of the kink scale. The latter approach was taken in previous
simulations \cite{GR}\cite{BD}\cite{AFG}. This solution is not without
flaws either, as rapid fluctuations can still appear as kinks. We
prefer to count kinks in the following way: at a particular time we
first find all zero-crossings. To test the legitimacy of a given
zero-crossing we check for zero-crossings one kink scale (8 lattice
units) to its right and to its left. If no zero-crossings are found,
we count it as a kink, otherwise not.

The number of kinks is plotted against $\beta$ in Fig. III. Above
$\beta\sim 6$, the averaged field method and our method for counting
kinks agree. Moreover, in this (low temperature) range, the dilute gas
expression for the kink number (\ref{dgnum}) is in excellent agreement
with the data. At elevated temperatures, there is a clear disagreement
between the two methods of counting kinks. The average field technique
has the number of kinks monotonically increasing with temperature;
whereas, in accord with intuition and the behavior of $P[{\phi}]$, the
second technique clearly shows a reduction in the kink number at
higher temperatures. Moreover, in this temperature regime the number of
kinks computed with the smoothing method depends strongly on the
smoothing scale. We conclude that for $\beta < 6$, the number of kinks
cannot be extracted with any confidence from the smoothing method.
Unfortunately, this is precisely the temperature regime explored in
previous simulations.

The correlation length $\lambda$ is plotted against $\beta$ in Fig.
IV. For $\beta > 6$, the WKB prediction (\ref{cl}) holds, whereas
for $\beta < 5$, where the wave function overlaps are not negligible,
the double Gaussian approximation is valid. Fortunately, there are no
ambiguities here with regard to measurements at higher temperatures.

As a consequence of the above results, we conclude that the dilute
gas/WKB approximation is excellent for $\beta > 6$ with no further phonon
dressing of the bare kink energy beyond that already included in
(\ref{dgnum}) and (\ref{CL}) at these low temperatures. At higher
temperatures, the WKB analysis fails, though theoretical progress is
possible with the double Gaussian technique.

We thank M. Alford, S. Chen, F. Cooper, G. D. Doolen, H.
Feldman, R. Gupta, R. Mainieri, M. Mattis, E. Mottola, W. H. Zurek,
and, especially, A. R. Bishop for encouragement and helpful
discussions. We are indebted to Alex Kovner for his numerous insights
and suggestions. This work was supported by the U. S. Department of
Energy at Los Alamos National Laboratory and by the Air Force Office
of Scientific Research. Numerical simulations were performed on the
CM-200 at the Advanced Computing Laboratory at Los Alamos National
Laboratory and the CM-2 at the Northeast Parallel Achitecture Center
at Syracuse University.

\section*{Figure Captions}

Fig. I: The classical distribution function $P[{\phi}]$ given by the
simulation (solid line) and the distribution $\Psi_0^{~2}$ from the
double Gaussian approximation (dashed line) plotted against
$\bar{\phi}$ for $\beta=2$. The potential
$V=-(1-\bar{\phi}^2/(2\beta^2))\bar{\phi}^2/(2\beta^2)$ of the
equivalent quantum problem (\ref{ham}) is also shown.

\noindent Figure II: Field configurations, from top to bottom, at
$\beta=2$, $\beta=4$, and $\beta=8$. Only a 1000 lattice unit sample
of the total lattice size of 16384 is shown.

\noindent Figure III: Total number of kinks and anti-kinks as a
function of $\beta$. Squares denote counts with a smoothed field
(smoothing scale=8 lattice units) definition of kinks, triangles for the
zero-crossing count method explained in the text, and the solid line
is a fit to (\ref{dgnum}). Where not shown, error bars are of the
order of the sizes of the symbols. The smoothed field result is in
agreement with the results of Ref. \cite{AFG} for $\b<5$.

\noindent Figure IV: Field correlation length $\lambda$ as a function
of $\beta$. The short-dashed line is the Gaussian prediction, whereas
the long-dashed line is the WKB result (\ref{CL}).


\begin{thebibliography}{50}

\bibitem{SS} See, e.g., A. Seeger and P. Schiller, in {\em
Physical Acoustics, Vol. III} edited by W. P. Mason (Academic, New
York, 1966).

\bibitem{CM} {\em Solitons and Condensed Matter Physics}, Proceedings
of the Symposium on Nonlinear Structure and Dynamics in Condensed
Matter, edited by A. R. Bishop and T. Schneider (Springer-Verlag, New
York, 1978).

\bibitem{BKT} A. R. Bishop, J. A. Krumhansl, and S. E. Trullinger,
Physica D {\bf 1}, 1 (1980).

\bibitem{SR} V. A. Kuzmin, V. A. Rubakov, and M. E. Shaposhnikov,
Phys. Lett. {\bf 155B}, 36 (1985).

\bibitem{KT} See, e.g., {\em The Early Universe} by E. W. Kolb and M.
S. Turner (Addison-Wesley, New York, 1990).

\bibitem{KS} J. A. Krumhansl and J. R. Schrieffer,
Phys. Rev. B {\bf 11}, 3535 (1975).

\bibitem{CKBT} J. F. Currie, J. A. Krumhansl, A. R. Bishop,
and S. E. Trullinger,
Phys. Rev. B {\bf 22}, 477 (1980).

\bibitem{SSF} D. J. Scalapino, M. Sears, and R. A. Ferrell,
Phys. Rev. B {\bf 6}, 3409 (1972).

\bibitem{KBKS} T. R. Koehler, A. R. Bishop, J. A. Krumhansl, and J. R.
Schrieffer,
Solid State Commun. {\bf 17}, 1515 (1975).

\bibitem{GR} D. Yu. Grigoriev and V. A. Rubakov
Nuc. Phys. {\bf B299}, 67 (1988).

\bibitem{BD}  A. I. Bochkarev and Ph. de Forcrand,
Phys. Rev. Lett. {\bf 63}, 2337 (1989).

\bibitem{AFG} M. Alford, H. Feldman, and M. Gleiser,
Phys. Rev. Lett. {\bf 68}, 1645 (1992).

\bibitem{SSB} E. Stoll, T. Schneider, and A. R. Bishop,
Phys. Rev. Lett. {\bf 42}, 937 (1979).

\bibitem{ScS} c.f., T. Schneider and E. Stoll,
in Ref. \cite{CM}.

\bibitem{BK} A. R. Bishop and J. A. Krumhansl,
Phys. Rev. B {\bf 12}, 2824 (1975).

\bibitem{AHK} F. J. Alexander, S. Habib, and A. Kovner (to be
submitted to Phys. Rev. E)

\bibitem{GA} See, e.g., P. M. Stevenson,
Phys. Rev. D {\bf 30}, 1712 (1984).

\bibitem{ADB} A. D. Bruce,
Adv. Phys. {\bf 29}, 111 (1980).

\bibitem{ARB} A. R. Bishop, in {\em Lattice Dynamics}, Proceedings of
the International Conference on Lattice Dynamics, Paris, 1977, edited
by M. Balkanski (Flammarion Sciences, Paris, 1978).
\end{thebibliography}
\end{document}